\documentclass[ASNA,twocolumn]{USG} 
\usepackage{anyfontsize} %

\usepackage{colortbl}   
\usepackage{xcolor}     

\usepackage{steinmetz}
\graphicspath{{./images/}}
\specialissue{Morphologic Hard and Soft Tissue Changes following Tooth
Extraction in Different Maxilla and Mandible Anatomic Areas: Current
Knowledge, Prognostic Factors Analysis, and Clinical Management}

\articletype{RESEARCH ARTICLE}%
\subarticletype{Particle Technology and Fluidization}

\journal{Advanced Optical Materials}
\volume{0}
\copyyear{Year}
\startpage{1}
\articledoi{10.1002/0000}

\begin{document}
\title{Excited-state lifetime and decay mechanisms in erbium-doped ZBLAN using Variable Excited-State dual-Pump Analysis}
\author[1,2]{Clement Chan}
\author[1]{Thomas J. de Prinse}
\author[1,3]{Jillian E. Moffatt}
\author[1,3]{Nigel A. Spooner}
\author[1,2]{David J. Ottaway}
\author[1,2]{Ori Henderson-Sapir}

\authormark{CHAN \textsc{et al.}}
\titlemark{Excited-state lifetime and decay mechanisms in Erbium using Variable Excited-State dual-Pump Analysis}

\address[1]{\orgdiv{Institute for Photonics, Advanced Sensing and Quantum Technologies (IPAS-QT), School of Physics, Chemistry and Earth Sciences, }\orgname{Adelaide University, }%
\orgaddress{\state{South Australia 5005, }\country{Australia}}}

\address[2]{\orgname{ARC Centre of Excellence for Gravitational Wave Discovery (OzGrav), }%
\orgaddress{\city{Hawthorn, }\state{Victoria 3122, }\country{Australia}}}

\address[3]{\orgname{CRC for Optimizing Resource Extraction, }%
\orgaddress{\street{PO Box 403, }\city{Kenmore, }\state{Queensland 4069, }\country{Australia}}}

\corres{Clement Chan (\email{clement.chan@adelaide.edu.au}) ~|~ Thomas J. de Prinse (\email{thomas.deprinse@adelaide.edu.au}) ~|~ Jillian E. Moffatt (\email{jillian.moffatt@adelaide.edu.au}) ~|~ Nigel A. Spooner (\email{nigel.spooner@adelaide.edu.au}) ~|~ David J. Ottaway (\email{david.ottaway@adelaide.edu.au}) ~|~ Ori Henderson-Sapir (\email{ori.henderson-sapir@adelaide.edu.au})}


\presentaddress{School of Physical Sciences, Adelaide University, Adelaide, South Australia}


\keywords{Lifetime technique | ZBLAN | Spectroscopy | Erbium }


\abstract[ABSTRACT]{This work presents a new technique for measuring excited-state lifetimes of rare-earth ions.  This technique is particularly powerful for states that are very short lived or have a low photon yield because their de-excitation pathways are dominated by non-radiative processes. The technique uses dual optical parametric oscillators, whose pulses are separated by a variable delay. It reconstructs the excited-state decay curve from a series of delayed fluorescence intensity spectra rather than directly recording the fluorescence decay as done in conventional time-resolved laser fluorescence spectroscopy. The power of the technique is demonstrated by measuring the lifetime of the short-lived $^4I_{9/2}$ level in Er$^{3+}$-doped ZBLAN, which is well-known for its low photon yield. The technique enables accurate measurement of its lifetime and suggests a new depopulating mechanism involving the $^4I_{9/2}$ level in ZBLAN glass. Its accuracy is validated by measuring the lifetime of the $^4F_{9/2}$ state achieving excellent agreement with conventional time-resolved fluorescence spectroscopy measurements. The approach is well suited for measuring lifetimes ranging from the sub-microsecond to the millisecond range. It provides a versatile platform for investigating excited-state dynamics and energy-transfer processes in RE-doped materials.}

\abbr{5-FU, 5-fluorouracil; CFD, computational fluid dynamics; CH, channel; EFS, event-free survival; GBM, glioblastoma multiforme; OS, overall survival; PFS, progression-free
survival; SD, standard deviation.}

\contributed{Hao Zhang and Pengyue D. Guo contributed equally to this study.}

\dedicated{Dedicated to Srivivasa Ramanujano on the occasion of his 125th birth anniversary.}

\copyright{Copyright Statement
\\[5pt]
  ©  2024 XXXX}


\maketitle


\section{Introduction}\label{sec1}

Trivalent rare-earth (RE) ions, including Er$^{3+}$, Yb$^{3+}$, Tm$^{3+}$, Pr$^{3+}$, and Nd$^{3+}$, doped into a variety of hosts have been extensively investigated because their optical properties make them suitable for numerous applications such as solar cells, optical sensors, telecommunication, semiconductor physics or optical amplifiers and laser sources \cite{Dwivedi_Respec_2014,pavesi_2004_REin_SemiCon,Krankel_REspec_LaserApplication}.

Knowledge of the excited-state lifetimes in RE-doped glasses is critical for understanding the population dynamics of laser systems. For the purpose of measuring RE lifetimes, conventional techniques such as frequency-domain modulation \cite{ES_Magden_LTTech_2018} and more commonly time-resolved laser fluorescence spectroscopy (TRLFS) are used \cite{GAFT_1998_LTTech,schnabel2000_LTTech_TRLFS,hannaford1983_LTTech_TRLFS}. Excited-state lifetime measurements provide insight into ion dynamics and their fundamental spectroscopic properties. Their dependence on the local environment makes them sensitive to both the host glass composition and the dopant concentration, thereby  influencing the rates of radiative and non-radiative transitions.
Spectroscopic investigations often involve calculating the Judd-Ofelt parameters which are important in quantifying how the host material affects the radiative absorption and emission properties of the RE ions \cite{reisfeld2007radiative_JUDDo, Jackson_2022_REspec_InF3_ZBLAN}. Understanding the non-radiative transitions affecting rare-earth (RE) species is essential, as these transitions are primarily driven by multiphonon decay and energy transfer between neighboring ions. Furthermore, the rate of multiphonon relaxation is governed by the electron-phonon coupling strength and the phonon energy of the host matrix~\cite{layne1977multiphonon}. According to Förster's theory \cite{ET_Forster_Energymigration_translated}, energy transfer occurs through electric dipole–dipole interactions between neighboring ions, where an excited donor transfers energy non-radiatively to a nearby acceptor. This mechanism can lead to energy migration between identical ions called donor–donor interactions \cite{ET_Watts_1972_diffusion}, or between different ions referred to as donor–acceptor interactions. The combination of radiative and non-radiative transitions dictates the lifetime of a given excited state. TRLFS is a standard approach for quantifying such lifetimes \cite{Ososkov_InF_2023}, investigating the impact of energy transfer processes \cite{Nkonde_CoDopedRE_2018,CoDopedRE_Zhao_2025_ErDy} and deepening the understanding of donor-donor and donor-acceptor interactions \cite{ET_Dexter_1953_OriginialTranslated,ET_Forster_Energymigration_translated}.

Conventional TRLFS measurements become increasingly challenging for short-lived excited states exhibiting low brightness, where the temporal resolution is limited by the excitation pulse duration and detector response. This creates a fundamental trade-off between response time and detection sensitivity as detectors with faster response times typically exhibit reduced sensitivity, while highly sensitive detectors often compromise temporal resolution due to slower response characteristics. This limitation is particularly relevant for heavily quenched RE levels, such as the $^4I_{9/2}$ state of Er$^{3+}$ in the fluoride glass ZBLAN, due to its relatively fast decay caused predominantly by non-radiative processes and hence low emission intensity. 
To the best of our knowledge, this is the first time $^4I_{9/2}$ state of Er$^{3+}$-doped ZBLAN decay dynamics is provided with high accuracy. This paper provides an in-depth methodology that could be applied to any RE lifetime and is particularly suited to transitions that are non-radiative decay dominated. 

In this work, we present a new approach for quantifying RE lifetimes based on fluorescence intensity generated by a temporally separated dual pump pulse excitation. 
The approach builds on a method used by de Prinse et al. in their nanoparticles upconversion (ETU) study \cite{Tom_2OPO_2021}
and we refer to the modified technique as the variable excited-state dual-pump analysis (VESPA) technique. Our approach indirectly reconstructs the time dependent excited-state population. Figure~\ref{energy level diagram for Innova} shows an energy level diagram for Er$^{3+}$-doped ZBLAN and will be used to illustrate how this technique works. An initial short pulse, shown in orange, excites the ion to the level of interest or state under test (SUT). A second, delayed pulse of comparable intensity with a different wavelength is applied. The second pulse is resonant with an excited-state transition from the SUT to a significantly longer-lived state whose predominant decay path is a single optical transition, see purple lines in Fig 1, and green for the decay path. The fluorescence emitted from the higher state can then be measured with ease. By varying the delay between the two excitations  and measuring the intensity of fluorescence form the longer-lived state, the decay dynamics of the SUT can be reconstructed using relatively slow, conventional detectors. This technique is therefore well suited to measuring rapid decays on the sub-microsecond timescale.

As a proof-of-concept we conducted an  experiment using bulk samples of Er$^{3+}$-doped ZBLAN (ZrF$_4$–BaF$_2$–LaF$_3$–AlF$_3$–NaF). We validated our approach against a conventional technique (TRLFS), using measurements of the lifetime of the $^4$F$_{9/2}$ level in $Er^{3+}$-doped ZBLAN. Then, we focused on the short-lived $^4$I$_{9/2}$ level of $Er^{3+}$ to demonstrate this new technique. The excellent agreement between the two independent measurement approaches demonstrates the reliability of the proposed VESPA method and highlights its potential for investigating short-lived excited states in RE-doped media.
 
\section{Method}

 This technique differs from conventional approaches as it reconstructs the lifetime by combining relative measurements of the fluorescence from a series of spectra, rather than relying on a direct time-resolved decay measurement. In this section, we describe how the two pumps interact with the sample to generate a series of fluorescence spectra that can faithfully reconstruct the lifetimes of excited states.

 \begin{figure}[h]
    \centering
    \includegraphics[width=0.85\linewidth]{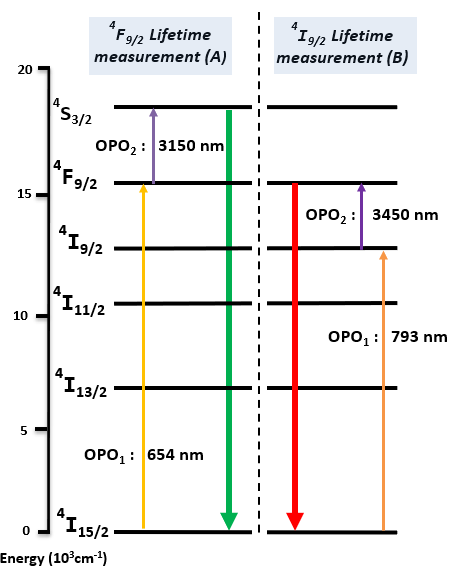}
    \caption{Schematic of Er$^{3+}$ energy levels illustrating the excitation wavelengths and the emission bands while using the VESPA technique for measuring the lifetime of the $^4$F$_{9/2}$ (left) and $^4$I$_{9/2}$ (right) energy levels at two different experiments.  The OPO$_1$ pulses used to establish a population in the state under test are shown in yellow. The OPO$_2$ pulses used to probe this population are shown in purple and the fluorescence that is measured is shown in green ($^4$S$_{3/2}$) and red ($^4$F$_{9/2}$).}
    \label{energy level diagram for Innova}
\end{figure}




\begin{figure}
    \centering
    \includegraphics[width=1\linewidth]{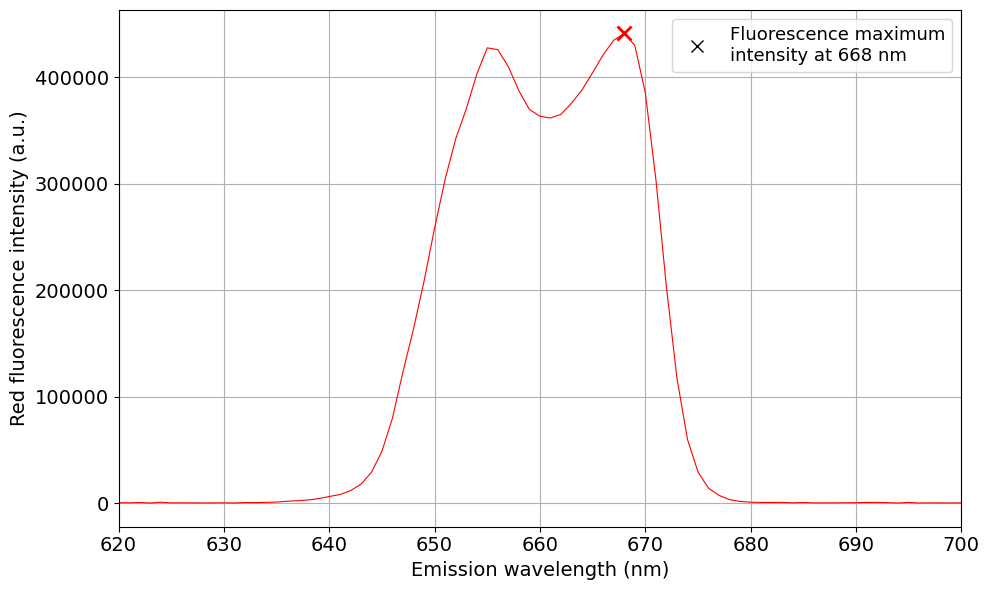}
    \caption{Red fluorescence from  $^4$F$_{9/2}$ level in Er$^{3+}$ doped ZBLAN. The maximum intensity is recorded at 668 nm, which is used to gauge the population dynamics of the $^4$I$_{9/2}$ level following the second, delayed pulse. This curve corresponds to a zero-delay measurement. The fluorescence spectra recorded at other delays exhibit an identical spectral profile.} 
    \label{serie of spectra for I9/2 LT}
\end{figure}

Figure~\ref{energy level diagram for Innova} illustrates the excitation pathway of the two optical parametric oscillator pulses (OPO) pumps noted OPO$_1$ and OPO$_2$. Following excitation, the fluorescence is measured. The red fluorescence centered at 650 nm and green fluorescence centered at 540 nm are recorded to extract the lifetimes of $^4$I$_{9/2}$ and $^4$F$_{9/2}$, respectively. 

In the case of the $^4$I$_{9/2}$ lifetime, OPO$_1$ first excites the sample and establishes a population of ions in that level. Next, OPO$_2$ is used to probe that population by promoting ions from $^4$I$_{9/2}$ to $^4$F$_{9/2}$. Initially, the delay between the two pulses is adjusted so that both OPO outputs propagate and overlap simultaneously at the sample. This condition defines the zero delay point corresponding to the first data point of the experiment. 
For a zero delay, the two pulses temporally overlap resulting in strong fluorescence, due to the direct excitation of the $^4$I$_{9/2}$ population to $^4$F$_{9/2}$ via excited-state-absorption, to a state with far higher photon yield and a longer lifetime.  Subsequently, the measurement is repeated for increased delay between the two OPO pulses. To improve the signal-to-noise ratio, the fluorescence is averaged over many excitation cycles (300 to 1000 ) whilst maintaining a constant delay between the pulses producing a spectrum, as shown in Figure~\ref{serie of spectra for I9/2 LT}.  The maximum fluorescence intensity is recorded at 668~nm, which is proportional to the $^4$I$_{9/2}$ population. When the delay time is increased, ions start to relax from the $^4$I$_{9/2}$ state, hence fewer ions are promoted to $^4$F$_{9/2}$ by OPO$_2$ resulting in a lower intensity of the red fluorescence. The delay time is increased until the red fluorescence disappears which is typically 50~$\mu s$ or about 5~lifetimes. The lifetime decay curve can be extracted by fitting the maximum intensity of the red fluorescence spectra as a function of the delay between the OPO$_2$ and OPO$_1$ pulses. This method can be reconfigured to measure  the~$^4$F$_{9/2}$ transition lifetime or other transitions by adjusting the OPOs excitation wavelengths to the correct energy levels.


\section{Set-up}

\begin{figure}[h]
    \centering
    \includegraphics[width=1\linewidth]{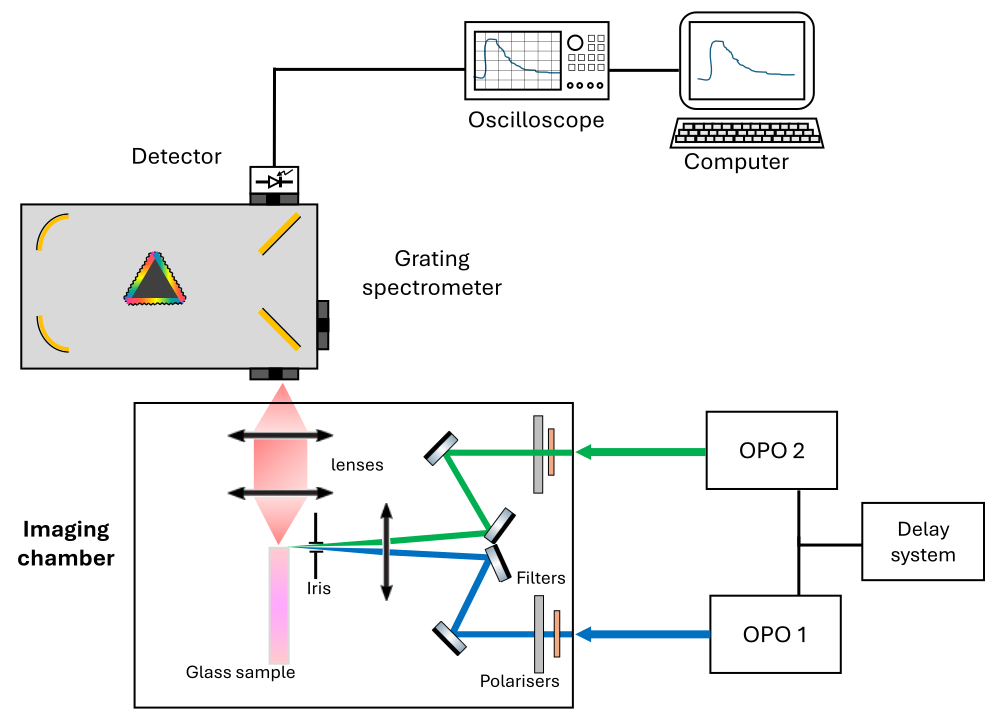}
    \caption{Schematic of our excited state measurement set-up}
    \label{Innova lifetime set-up}
\end{figure}

A schematic of our measurement technique is shown in Figure~\ref{Innova lifetime set-up}. OPOs are used as tunable excitation sources. The first pump, noted OPO$_1$ on Figure~\ref{Innova lifetime set-up} is an Opotek Radiant-X with a tunable wavelength range of 410~nm to 2500~nm. It is pumped by the third harmonic (355~nm) of a Q-switched 1064~nm Nd:YAG laser, with a pulse duration of 7~ns and repetition rate of 10~Hz. The second pump referred as OPO$_2$ on Figure~\ref{Innova lifetime set-up}, is an Ekspla NT377 mid-infrared (MIR) OPA, with a pulse duration of 6ns providing tunable excitation from 2500 to 4400~nm at a repetition rate of 10~Hz. Like OPO$_1$, it is also pumped with the triple harmonic (355~nm) of a Q-switched 1064~nm Nd:YAG laser. The delay system adjusts the trigger timing of one OPO relative to the other, enabling control of the pulse separation with a 1~ns resolution, significantly finer temporal resolution compared with the SUTs lifetimes. The entire system is controlled using an in-house designed control software and use a low jitter pulse trigger system

Pulse energies are wavelength dependent and can be controlled by external polarizers prior to the sample in the imaging chamber. The imaging chamber can accommodate powder samples as well as bulk samples of dimension of about 3~cm$^2$. Following excitation of the sample, fluorescence emission is recorded using two spectrometers covering complementary wavelength regions. The emission range of 300-1000~nm is measured using a Princeton Instruments Acton SpectraPro SP-2300 spectrometer coupled with a PIXIS 100 silicon CCD.

\section{Results}

\subsection{$^4$F$_{9/2}$ decay characteristics}

To validate the proposed approach the first state to be tested is $^4$F$_{9/2}$ in Er$^{3+}$-doped ZBLAN. The intrinsic lifetime of the $^4$F$_{9/2}$ level in ZBLAN glass varies in the literature ranging between 52~$\mu$s - 177~$\mu$s \cite{Zou_ErTm_1995,bogdanov1999energy_thesis,Ori_thesis,Skrzypzak_2011_Er_LT}. The $^4$F$_{9/2}$ lifetime is significantly influenced by dopant-dependent energy transfer process which can corrupt lifetime measurements if a significant population in the $^4$I$_{11/2}$ level is established during the measurement~\cite{REspec_ORI_NewCR_2016}. Our method avoids this issue.

\begin{figure}[h]
    \centering
   
    \includegraphics[width=0.85\linewidth]{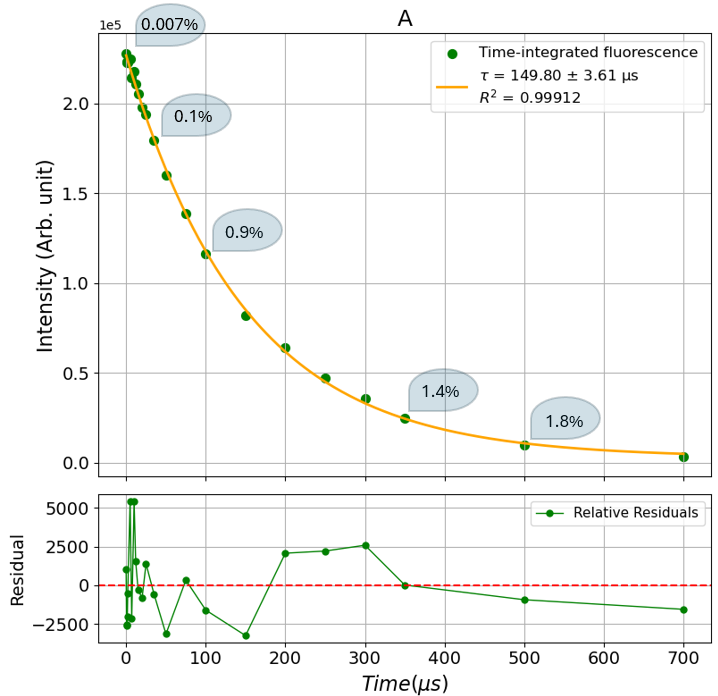}

    \vspace{0.5cm}
    
    \includegraphics[width=0.85\linewidth]{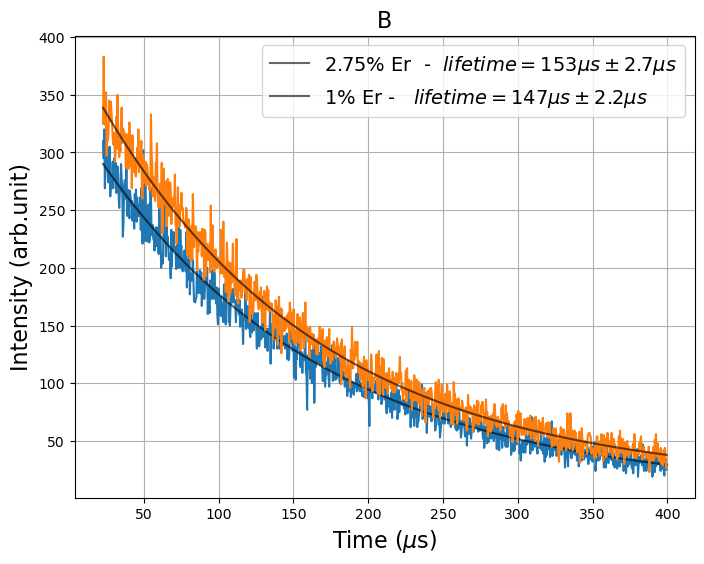}
    
    \caption{ Intrinsic lifetime of Er$^{3+}$-doped ZBLAN $^4$F$_{9/2}$ state. (A) Lifetime measured using the novel approach corresponding to 1\% doping concentration, OPO$_1$ excitation is at 654~nm, delayed with OPO$_2$ at 3150~nm. The call-outs display the relative uncertainty in the measured fluorescence intensity value. (B) Two decay curves at different doping concentrations (1\% and 2.75\% ), measured using the conventional time-resolved technique.}
    \label{F9/2 as validation}
\end{figure}

 In the present study, both VESPA and conventional TRLFS experimental approaches are used to provide a comprehensive analysis of the $^4$F$_{9/2}$ lifetime. Illustrated by (A) on Figure~\ref{F9/2 as validation}, VESPA yielded a  $^4$F$_{9/2}$ lifetime of 150~$\mu$s for a Er$^{3+}$ sample of 1~mol.\% doping concentration. In comparison, TRLFS, shown by (B), produced the fitted lifetimes of 153 $\mu$s and 147~$\mu$s for samples with 2.75~mol.\% and 1~mol.\% concentration, respectively.

The uncertainty calculated from the covariance matrix is shown in Figure~\ref{F9/2 as validation}(A), residuals. TRLFS technique (B), has a fitted lifetime uncertainty of 2.2~$\mu$s, whereas VESPA approach (A) gives an uncertainty of 4~$\mu$s for a 1mol.\% sample. 

Based on the covariance matrix, TRLFS technique provides a lower statistical uncertainty despite the presence of apparent noise. However, these uncertainty values should be interpreted with caution, as the noise contributions differ between the two measurement approaches. For TRLFS, the noise level is directly visible in the fluorescence decay curve and is of similar magnitude across the decay curve, because it is mainly determined by the detector sensitivity. In contrast, the noise level of the VESPA approach is less apparent, as each fluorescence intensity value is extracted from an integrated spectrum. As the delay increases, the measured fluorescence intensity approaches the noise floor, resulting in a larger relative uncertainty in the extracted intensity values. The relative uncertainty of selected data points is illustrated in Figure~\ref{F9/2 as validation}A and was determined by repeating the same measurement twice. These numbers increase from 0.007\% for the initial intensity value to 1.8\% at the end, demonstrating the increasing influence of noise as the fluorescence signal decays. These uncertainties are too small to be seen by error bars on (A). 

The measured lifetimes remain in excellent agreement between the two independent approaches, demonstrating the accuracy of the proposed method. These results resolve the uncertainty in the literature surrounding the lifetime of the $^4$F$_{9/2}$ level in Er$^{3+}$-doped ZBLAN for doping concentrations below 3~mol.\%, allowing its value to be confidently established as 150~$\mu$s for future studies. 

\subsection{$^4$I$_{9/2}$ decay characteristics }

\begin{figure}[htb]
    \centering
  
    \includegraphics[width=0.85\linewidth]{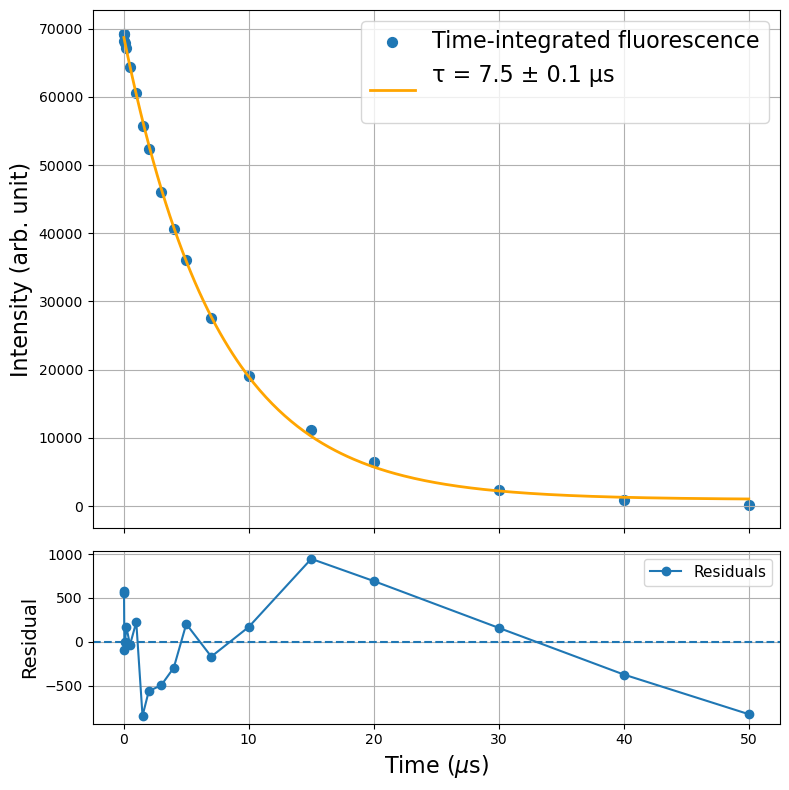}

    \caption{Fluorescent measurements and single exponential fit for measuring the lifetime of the $^4$I$_{9/2}$ state in 0.6~mol.\% Er$^{3+}$-doped ZBLAN. OPO$_1$ is set at 793~nm and OPO$_2$ at 3450~nm. The resulting red fluorescence intensity from the higher excited-state $^4F_{9/2}$ is collected corresponding to the data points shown. The residuals from the single exponential are also shown}
    \label{I9/2 tau}
    
\end{figure}

The fluorescence decay from the $^4$I$_{9/2}$ state was expected to follow a single-exponential law given its short lifetime and the expectation that energy transfer processes would not be significant due to this short timescale. Our initial estimate of lifetime of the $^4$I$_{9/2}$ state was obtained using a single exponential fit shown in Fig \ref{I9/2 tau}. The residuals are also shown to assess the quality of the fit. The structure observed in the residuals indicates that the fluorescence decay is not perfectly described by a single-exponential model. 

We will now show that better fits can be obtained when a double exponential or an upconversion process is added. This could only be observed because of the high signal-to-noise of our method.

\subsubsection{Double-Exponential Decay Model}
\vspace{0.2cm}

\begin{figure}[htb]
    \centering
  
    \includegraphics[width=1\linewidth]{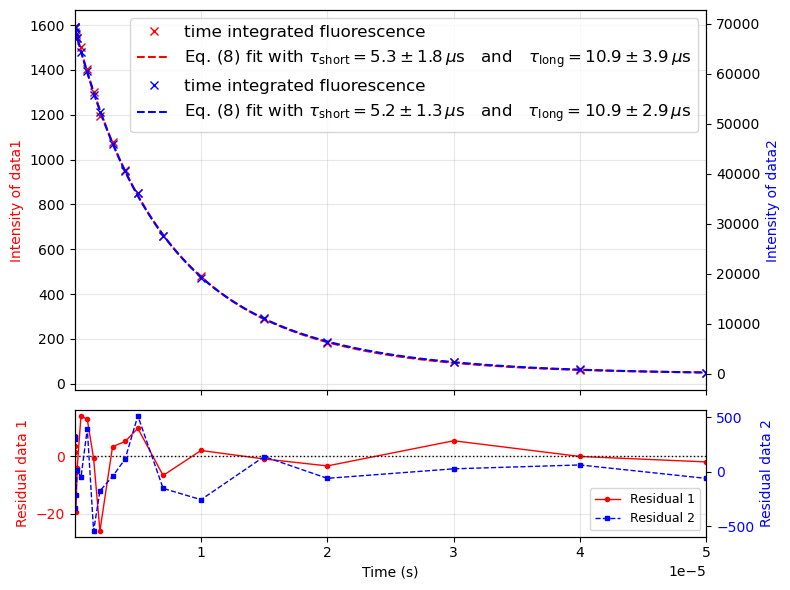}

    \caption{$^4$I$_{9/2}$ decay in 0.6~mol.\% Er$^{3+}$:ZBLAN, measured by the VESPA method. Two sets of time-integrated fluorescence are then fitted with an Eq.~\ref{N2EXP} that characterize the evolution of 2 independent population $\tau_{short}$ and $\tau_{long}$ }
    
    \label{2EXP I9/2}
    
\end{figure}

 The fluorescence decay of $^4$I$_{9/2}$ is fitted using a double-exponential model as shown by Figure~\ref{2EXP I9/2}. This approach assumes two independently decaying excited-state populations, each characterized by a distinct lifetime, as shown below :

\begin{equation}
\frac{dN_1}{dt}=-\frac{N_1}{\tau_1}, \qquad
\frac{dN_2}{dt}=-\frac{N_2}{\tau_2}.
\label{2EXP}
\end{equation}

Their analytical solutions yield the following expression for the total fluorescence intensity:

\begin{equation}
N(t)=A_1\exp(-t/\tau_{short})+A_2\exp(-t/\tau_{long})+B,
\label{N2EXP}
\end{equation}

where \(A_1\) and \(A_2\) are the amplitudes of the two decay components, \(\tau_1\) is the short lifetime and \(\tau_2\) is the long lifetime. This model neglects nonlinear depopulation mechanisms associated with excitation transfer between ions, but instead describes the fluorescence decay as the superposition of two exponential contributions. This fit and its residual are represented below on Figure~\ref{2EXP I9/2}. The difference between the Data 1 and Data 2 is the averaging time of the VESPA collected fluorescence which is 30~s and 180~s, respectively. Note that both datasets exhibit consistent fitted time constants, although Data 1 (red) has slightly higher uncertainties. This is explained by the shorter averaging time of the integrated fluorescence, resulting in a lower signal to noise ratio. The decay consists of two contributions: A rapid component with a fitted time constant of approximately (5~$\mu$s), and a slower component with a time constant of approximately (11~$\mu$s). The magnitude of the residuals are $\times30$ lower compared with those from the single-exponential function fit and are much more consistent with random processes. Therefore the data is shown to be more consistent with a mathematical model that contains two time constants.

\subsubsection{ Upconversion-based decay model} 
\vspace{0.2cm}

As an alternative to the double exponential model, the fluorescence decay was also fitted an upconversion model. It is well known that upconversion processes are described by the following excited state population equation \cite{ET_Glolding_Jackon_2000_Diffusion}:

\begin{equation}
    \frac{dN}{dt} = -\frac{N}{\tau} - 2W_{ETU}N^2
    \label{Eq1}
\end{equation}

where $\tau$ is the intrinsic lifetime of $^4$I$_{9/2}$ and W$_{ETU}$ is the upconversion parameter relating to the process. With $N(\tau = 0) = N_{I_{9/2}}$ state population density. This differential equation has an analytic solution of:


\begin{equation}
    N(t) = \frac{1/\tau \: N_{I_{9/2}}\: exp(-t/\tau)}{1/\tau + 2W_{ETU}N_{I_{9/2}}\: [1 - exp(-t/\tau)]}
    \label{Eq2}
\end{equation}

\begin{figure}[htb]
    \centering
  
    \includegraphics[width=1.05\linewidth]{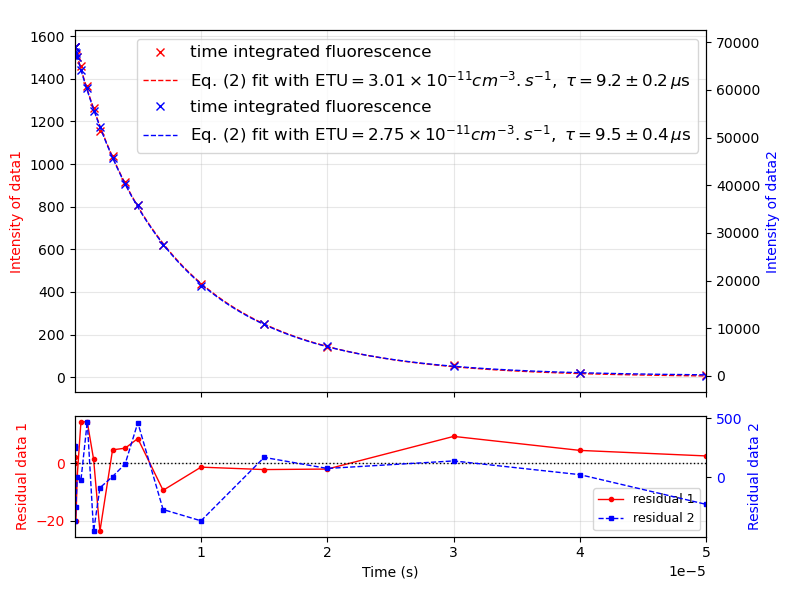}

    \caption{$^4$I$_{9/2}$ decay in 0.6~mol.\% Er$^{3+}$:ZBLAN, measured by the VESPA method utilizing two pumps. Two sets of time-integrated fluorescence are then fitted with Eq.~\ref{Eq2} that characterize W$_{ETU}$ and the intrinsic lifetime $\tau$}
    \label{I9/2 etu and tau}
    
\end{figure}

To fit this data from Figure~\ref{I9/2 tau} we used the following formula with free parameters $A$, $B$ and $\tau$ to the fluorescent data:

\begin{equation}
    J(t) = \frac{Aexp(-t/\tau)}{1/\tau + B[1 - exp(-t/\tau)]}
    \label{Eq2a}
\end{equation}

This fit and its residuals are shown for two sets of data using the same sample of 0.6~mol.\% Er$^{3+}$ -doped Fig~\ref{I9/2 etu and tau}. The structure of the residuals is very similar with the double exponential fit. The comparable magnitude obtained from both models indicates that the measurement precision is insufficient to discriminate between the ETU and double-exponential decay models, this will be further discuss in the next section. 

The underlying lifetime is denoted by $\tau$, while the fitted amplitude $A$ is of secondary importance, as it primarily represents the fluorescence collection efficiency. Although the strength of the ETU process is represented by $B = 2W_{ETU}N_0$ extracting its value requires an estimate of the initial excited-state population of the $^4$I$_{9/2}$ state, which is obtained as follows.

The Beer--Lambert law describes the attenuation of light from $OPO_1$ as it propagates through the sample. 

\begin{equation}
I(x) = I(0)\exp\left(-\sigma_{\mathrm{abs}}N_g x\right)
\end{equation}

Where $I(0)$ the incident intensity, N$_g$ is the dopant concentration of the sample under investigation and $\sigma_{abs}$ the absorption cross-section of $^4I_{15/2} \rightarrow {^4}I_{9/2}$ transition in Er$^{3+}$. 

When propagation distance is small i.e. $x \rightarrow \Delta x$, the exponential can be approximated : 

\begin{equation}
I(\Delta x)
\approx
I(0)\left(1-\sigma_{\mathrm{abs}}N_g\Delta x\right),
\end{equation}

Dividing the absorbed intensity ($I_{abs} = I(0)\sigma_{abs}N_g\Delta x$ ) by the photon energy $h\nu$ and $\Delta x$ gives the value of $N_{I_{9/2}}$ assuming that each absorbed photon produces one excitation to the $^4$I$_{9/2}$ level, the initial population density can therefore be approximated as

\begin{equation}
N_{I_{9/2}} = \frac{\sigma_{\mathrm{abs}}N_gF(0)}{h\nu}.
\label{N2_initial}
\end{equation}

Where $F(0)$ is the pump fluence. The absorption cross-section $\sigma_{abs}$ at 793~nm can be found in the literature to be 6.26 x 10 $^{-22}$ cm$^{2}$ \cite{Pollnau_REspec_1998}, the Er$^{3+}$ dopant concentration $N_g$ in 0.6~mol.\%  Er$^{3+}$ -doped ZBLAN glass is 9.6 x 10$^{-21}$ cm$^{-3}$. The fluence is the energy per surface area, calculated using the pulse energy of the OPO (E = 0.5~mJ) and the focused area on the sample (S = 2 x 10$^{-1}$ cm$^{-2}$). Using the quantities listed above, the corresponding concentration is calculated to be $N_{I_{9/2}}$ = 6.0 x 10$^{-14}$ cm$^{-3}$. 

From Figure~\ref{I9/2 etu and tau}, the fits of data$_1$ (red) and data$_2$ (blue) exhibit good agreement using a different averaging time. Data$_1$ yields a lifetime $\tau = 9.2 \pm 0.2  \mu s $ and $W_{ETU}N_0 = 18323 $. Using the above analysis this yields an upconversion parameter W$_{ETU} = 3.01 \times 10^{-11}$ cm$^{3}$s$^{-1}$. Data$_2$ returns consistent results with $W_{ETU}N_0 = 16527 $ and W$_{ETU} = 2.75 \times 10^{-11}$ cm$^{3}$s$^{-1}$. Bogdanov's quantified this lifetime in \cite{bogdanov1999energy_thesis} using the time-resolved conventional technique and reported a value of 9.0~$\mu$s corresponding to 1.6~mol.\% . Additionally, Skrzypczak et al. also characterize the $^4$I$_{9/2}$ decay in ZBLAN and reported 7~$\mu$s for 1~mol.\% \cite{Skrzypzak_2011_Er_LT}. The reported values from the literature are consistent with our result. It is worth noting that neither of these references include the relevant decay curves and as a result it is not possible to comment on the signal to noise of their measurements and whether it would have been possible for them to observe this upconversion process.

\section{Discussion and future work}

We have shown that using the data obtained by the VESPA method both the upconversion and double exponential models provide suitable fits to the $^4$I$_{9/2}$ decay characteristics. The residual structures obtained from both models are similar, making it difficult to distinguish between them. In the presence of upconversion, the initial decay exhibits a quadratic dependence on the excited-state population, whereas the double-exponential model describes the decay as two independent exponential contributions, each characterized by a linear decay rate. At present, the early-time decay data do not provide sufficient evidence to determine which mechanism governs, so both the ETU and the double exponential model will be discussed as candidates. 

In the case of a double-exponential decay, one possible interpretation is the presence of distinct Er$^{3+}$ ion populations exhibiting different relaxation rates. In this scenario, the fast component ($\tau_{\mathrm{fast}}\approx 5~\mu$s) could originate from a subpopulation experiencing faster relaxation due to differences in their local environments, such as quenching. These differences may affect both radiative and non-radiative relaxation pathways. The slower component ($\tau_{\mathrm{slow}}\approx 11~\mu$s) would then correspond to ions experiencing a lower overall relaxation rate. Similar considerations regarding quenching of ion populations and their dynamics have been discussed by Agazzi et al. \cite{Pollnau_fastQuenching} for Er$^{3+}$ in amorphous Al$_2$O$_3$, exhibiting quenching of subpopulation of excited ions due to undesired impurities and defects. In their work they showed that the quenching effect could only be observed under short pulse excitation regime due to negligible emission of the quenched ions under quasi-continuous wave excitation. 
Gorjan et al. \cite{gorjan2011role_ETU_Quenching_Er} investigated the role of interionic processes in Er$^{3+}$-doped fluoride glass fibers and highlighted the influence of the local ion environment, including possible clustering. The strength of the interactions was highly sensitive to the separation between neighboring Er$^{3+}$ ions, resulting in distinct effective relaxation rates within the glass matrix. Numerical modeling achieved the closest agreement with data when two separate excited ion classes were included

To the best of our knowledge, the upconversion process from the $^4$I$_{9/2}$ level has not been previously reported. This is possibly because the techniques used to study the decay of this state lacked sufficiently high signal to noise ratio. 

\begin{figure}[htb]
    \centering
  
    \includegraphics[width=0.85\linewidth]{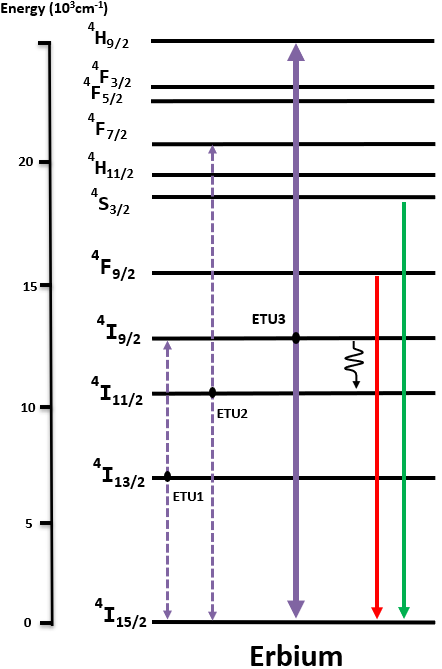}

    \caption{Energy levels of erbium ion illustrating upconversion processes. ETU$_2$ and ETU$_1$ are well-established upconversions in erbium depleting population from $^4$I$_{11/2}$ and $^4$I$_{13/2}$, respectively \cite{Jillian.Moffat_REspec_2011_ET_Er}. ETU$_3$ is the proposed new upconversion from $^4$I$_{9/2}$. }
    
    \label{I9/2 energy level of ETU}
    
\end{figure}

Examining the extended energy spectrum of erbium shown in Figure~\ref{I9/2 energy level of ETU} shows that the ($^4$I$_{9/2}$, $^4$I$_{9/2}$) $\rightarrow$ ($^4$H$_{9/2}$, $^4$I$_{15/2}$) is a strong, near-resonant upconversion candidate for this energy transfer process in Er$^{3+}$-doped ZBLAN. The reported energy positions of Er$^{3+}$ -doped ZBLAN of the $^4$I$_{9/2}$ and $^4$H$_{9/2}$ manifolds are 12,538 cm$^{-1}$ and 24,631 cm$^{-1}$, respectively \cite{Huang_REspec_ZBLAN_2001} and the maximum phonon energy of ZBLAN is 560 cm$^{-1}$. The energy mismatch between the $^4$I$_{9/2}$ $\rightarrow$ $^4$H$_{9/2}$ and $^4$I$_{9/2}$ $\rightarrow$ $^4$I$_{15/2}$ is less than the maximum phonon energy of ZBLAN meaning that energy transfer can occur with the creation of only a single phonon. This near-resonance condition makes the proposed ETU process highly likely under high population conditions. Fluorescence from $^4$H$_{9/2}$ is strongly quenched through non-radiative relaxation decaying down to $^4$S$_{9/2}$ resulting in green fluorescence. Consequently, it is difficult to distinguish this process in Er$^{3+}$ from ETU2 or an excited-state absorption (ESA) process.

It is important to acknowledge that the calculated values of ETU$_3$ from $^4$I$_{9/2}$ at 3.01 x 10$^{-11}$ cm$^{3}$s$^{-1}$ and 2.75 x 10$^{-11}$ cm$^{3}$s$^{-1}$ are higher by several orders of magnitude compared with any reported value of ETU$_2$ from $^4$I$_{11/2}$ and ETU$_1$ from $^4$I$_{13/2}$ at 0.16 x 10$^{-17}$ cm$^{3}$s$^{-1}$ and 1.3 x 10$^{-17}$ cm$^{3}$s$^{-1}$, respectively \cite{Maes_MID_laser_2019_refModel}. Note the transitions for ETU$_1$ and ETU$_2$ are shown on Figure~\ref{I9/2 energy level of ETU} for reference. Similarly to  ETU$_3$, both of these energy transfer processes are resonant processes with less than one maximum phonon energy involved. 


Beyond lifetime measurements, the flexibility of the two-OPO configuration enables selective excitation and probing of different energy levels within the same system, providing a powerful tool for investigating complex energy-transfer pathways in co-doped materials. Co-doping RE ions is widely used in laser physics to quench population at unwanted levels, or leverage the potential of energy transfer where ion populations are exploited to enhance laser performance \cite{Cantelar­_ErYb_2023}. 

\section{Conclusion}

A new approach for measuring RE excited-state lifetimes has been demonstrated using a variable
excited-state dual-pump analysis technique. The method was validated through comparison with conventional measurements of the $^4F_{9/2}$ level of Er$^{3+}$. The method also measured the lifetime of the short-lived $^4I_{9/2}$ level of Er$^{3+}$ -doped ZBLAN glass and revealed the presence of a previously unobserved process. The fast component of the new decay mechanism could be associated with a subpopulation of ions undergoing quenching interactions or could be attributed to a newly identified upconversion process. The root cause of this quenching will be investigated in future work.

The flexibility of the dual-OPO configuration described here, provides a powerful platform for investigating complex excited-state dynamics, including energy-transfer processes, cross-relaxation mechanisms, and poorly characterized cross-section.

\bmsubsection*{Author Contributions}

\textbf{Clement Chan}: investigation, data curation, writing -- original draft. \textbf{Thomas de Prinse}: methodology, investigation -- review \& editing. \textbf{Jillian Moffatt}: review.  \textbf{Nigel A. Spooner}: resources. \textbf{David Ottaway}: supervision, writing -- review \& editing. \textbf{Ori Henderson-Sapir}: conceptualization, investigation, supervision, writing -- review \& editing.

 \bmsubsection*{Acknowledgments}
 This research was funded by the Commonwealth Government Department of Defence, Next Generation Technologies Fund, Grand Challenge—Counter Improvised Threats (CIED) and CRC optimizing Resource Extraction (ORE) grants. We thank Stuart Jackson for valuable discussions. This work was performed in part at the Opto-Fab node of the Australian National Fabrication Facility supported by the Commonwealth and SA State Government. The authors acknowledge the support received from the Commonwealth of Australia through an Australian Research Council (ARC) Discovery Grant No. DP220102516 and Centre of Excellence Grant Nos. CE230100016, CE230100006

\bmsubsection*{Conflicts of Interest}

The authors declare no conflicts of interest.

\bmsubsection*{Data Availability Statement}

The data that support the findings of this study are available from the corresponding author upon reasonable request.

\bibliography{My_EndNote_Clement_corrected}




\end{document}